\documentclass[prl,twocolumn,showpacs,floatfix]{revtex4}
\usepackage{graphicx,amsfonts,amssymb,amsmath, hyperref}

\newif\ifhyper
\hypertrue
\ifhyper
\hypersetup{
     citecolor = {green},
   colorlinks = {true},                      urlcolor = {blue} }
\fi
\newcommand{\beq}{\begin{equation}}
\newcommand{\eeq}{\end{equation}}
\newcommand{\beqa}{\begin{eqnarray}}
\newcommand{\eeqa}{\end{eqnarray}}

\def\Longarrow{\protect\@lra}
\def\@lra{\relbar\joinrel\relbar\joinrel\relbar\joinrel          \relbar\joinrel\rightarrow}

\begin{document}

\title{Effective field theory for the SO($n$) bilinear-biquadratic spin chain%
}
\author{Hong-Hao Tu}
\author{Rom\'an Or\'us}
\affiliation{Max-Planck-Institut f\"ur Quantenoptik, Hans-Kopfermann-Str. 1, 85748
Garching, Germany}

\begin{abstract}
We present a low-energy effective field theory to describe the SO($n$)
bilinear-biquadratic spin chain. We start with $n=6$ and construct the
effective theory by using six Majorana fermions. After determining various
correlation functions we characterize the phases and establish the relation
between the effective theories for SO(6) and SO(5). Together with the known
results for $n=3$ and $4$, a reduction mechanism is proposed to understand
the ground state for arbitrary SO($n$). Also, we provide a generalization of the Lieb-Schultz-Mattis theorem for SO($n$).
The implications of our results for entanglement and correlation functions are discussed.
\end{abstract}

\pacs{75.10.Pq, 75.10.Jm, 03.65.Fd}
\author{}
\maketitle

\emph{Introduction.-} The study of quantum spin chains dates back to the
early days of quantum mechanics \cite{Bethe-1931}. With seemingly simple
Hamiltonians, quantum spin chains contain very rich physics and thus attract
a considerable interest. A prominent example is the bilinear-biquadratic
spin chain with SO($n$) symmetry,
\begin{equation}
H=\cos \theta \sum_{j}\sum_{a<b}L_{j}^{ab}L_{j+1}^{ab}+\sin \theta
\sum_{j}(\sum_{a<b}L_{j}^{ab}L_{j+1}^{ab})^{2}.  \label{eq-BB}
\end{equation}%
In the above expression $L^{ab}$ ($1\leq a<b\leq n$) are the generators of
SO($n$) in the $n$-dimensional vector representation, with the Casimir
operator normalized at every site as $\sum_{a<b}(L_{j}^{ab})^{2}=n-1$. Thus,
Eq.~(\ref{eq-BB}) represents a family of Hamiltonians parametrized
by $\theta \in \lbrack -\pi ,\pi ]$ and $n$. As an example, the case of $n=3$
corresponds to the well-known spin-$1$ bilinear-biquadratic model, of great
relevance in the context of Haldane's conjecture \cite{Haldane-1983}. Also,
for $n=4$ the model is equivalent to a symmetrically coupled spin-orbital
chain. Such spin-orbital models describe a family of transition metal oxide
compounds, where orbital degeneracy plays an important role in the magnetic
properties of the material \cite{Kugel-1973}. Quite remarkably, the phase
diagram of the model in Eq.~(\ref{eq-BB}) for these two cases has been established, and exhibits a rich variety of phases \cite%
{Laeuchli-2006,Itoi-2000,Azaria-2000}. However, the properties for $n\geq 5$
remain mostly unclear yet. For instance, it is unknown whether the behavior
in the large-$n$ limit is somehow similar to that of the models with smaller
$n$ or not.

In this context, the main contributions of this Letter are (i) to provide a generalization
of the Lieb-Schultz-Mattis theorem for SO($n$) and (ii) to present an
effective field theory describing the low-energy physics of the model in
Eq.~(\ref{eq-BB}) for arbitrary $n$ and $\theta \in \lbrack 0,\theta _{%
\mathrm{MPS}}]$, with $\theta _{\mathrm{MPS}}$ a special point to be
discussed later. This is relevant since it allows us to understand the
low-energy physics of the system in the considered parameter regime \emph{in
an exact way}. To achieve this goal, we start with $n=6$ and show that at a
given phase transition point $\theta _{\mathrm{R}}$ the system is critical
and is described by the SO(6)$_{1}$ Wess-Zumino-Novikov-Witten model
with central charge $c=3$. This, in turn, allows us to determine various
correlation functions characterizing the phases. Then, after establishing a
relationship between the effective theories for SO(6) and for SO(5) \cite%
{Alet-2010} and gathering some known results for $n=3$ and $4$, we propose a
reduction mechanism to understand the ground state of the SO($n$) Heisenberg
chain and suggest an effective field theory for Eq.~(\ref{eq-BB}%
) in the considered parameter regime.

\emph{Known results for the general SO($n$) case.-} Let us now comment
briefly on what is known about the model in Eq.~(\ref{eq-BB}) for arbitrary $%
n$. In Ref. \cite{Tu-2008} a phase diagram was conjectured based on the
existence of some exactly solvable points. For $n\geq 5$, this phase diagram
exhibits two remarkable features between the SO($n$) Heisenberg model $%
\theta _{\mathrm{H}}=0$ and the integrable SU($n$) Uimin-Lai-Sutherland model $\theta _{\mathrm{ULS}}=\tan ^{-1}\frac{1}{n-2}$ \cite%
{Sutherland-1975}. The first one is a parity effect in $n$; that is, the
physical properties of the system are sharply different if $n$ is even or
odd. This is best seen at the special point $\theta _{\mathrm{MPS}}=\tan
^{-1}\frac{1}{n}$, where the ground state of the model is exactly described
by a matrix product state (MPS) \cite
{Tu-2008,Affleck-1987,Kolezhuk-Scalapino}. This MPS is unique and translationally invariant for odd $n$, whereas it is twofold
degenerate and breaks translational symmetry for even $n$. The second
feature is the location of an integrable model at $\theta _{\mathrm{R}}=\tan
^{-1}\frac{n-4}{(n-2)^{2}}$ \cite{Reshetikhin-1983}, which sits between $%
\theta _{\mathrm{H}}$ and $\theta _{\mathrm{MPS}}$ for $n\geq 5$. This point
turns out also to be critical for all $n$, which has important implications.
For instance, the existence of this point implies that for $n$ $\geq 5$ the
MPS point $\theta _{\mathrm{MPS}}$ no longer captures the physics of the SO($%
n$) Heisenberg chain. For $n=5$ this feature is supported by an interesting
recent work \cite{Alet-2010}. Thus, the model in Eq.~(\ref{eq-BB}) for $n$ $%
\geq 5$ has a quite different phase diagram from the $n=3$ case,
where the Affleck-Kennedy-Lieb-Tasaki model\ at $\theta _{\mathrm{MPS}}$
qualitatively describes the properties of the spin-$1$ Heisenberg chain \cite%
{Affleck-1987}.

\emph{Generalized Lieb-Schultz-Mattis theorem.-} From a general perspective,
the parity effect has its roots in the difference of the SO($n$) vector
representation for odd and even $n$. More precisely, for \textit{even} $n$,
there exists an element $g\in $ SO($n$) such that $\exp (i\pi g)=-I_{n\times
n}$. The presence of this element enables a generalization of the
Lieb-Schultz-Mattis theorem \cite{LSM-1961}: Assuming that $|\Psi
\rangle $ is the unique ground state of the model in Eq.~(\ref{eq-BB}), the
\textquotedblleft twisted\textquotedblright\ state $|\Psi _{e}\rangle =\exp
(i\frac{\pi }{N}\sum_{j=1}^{N}jg_{j})|\Psi \rangle $ is not only orthogonal
to $|\Psi \rangle $ but also has a vanishing excitation energy in the
thermodynamic limit $N\rightarrow \infty $. This implies that the model with
\textit{even }$n$ has either gapless excitations or degenerate gapped ground
states with broken translational symmetry. On the contrary, the model with
\textit{odd }$n$ can have a unique gapped ground state.

\emph{Effective theory for $n=6$.-} Let us consider now the model in Eq.~(\ref%
{eq-BB}) for $n=6$, whose phase diagram is represented in Fig.~\ref%
{fig:SO(6)}. Since SO(6)$\simeq $SU(4), this model is equivalent to the
SU(4) spin chain with self-conjugate representation $\mathbf{6}$ in Ref.
\cite{Affleck-1991}. For $\theta \in \lbrack 0,\theta _{\mathrm{MPS}}]$, the
conjectured phase diagram in Ref. \cite{Affleck-1991} is in agreement with
Ref. \cite{Tu-2008} and the ground state at $\theta _{\mathrm{MPS}}$ is
identified as an extended valence-bond solid state with broken
charge-conjugation symmetry. Moreover, an effective theory was derived in
Ref. \cite{Affleck-1991} by using non-Abelian bosonization techniques. Here,
though, we use a different approach to obtain an effective theory, which
makes a clear connection with the SO(5) effective theory in Ref. \cite%
{Alet-2010} and enables a possible extension to general $n$. As expected,
this effective theory recovers the results from Ref. \cite{Affleck-1991}.

\begin{figure}[tbp]
\includegraphics[scale=0.23]{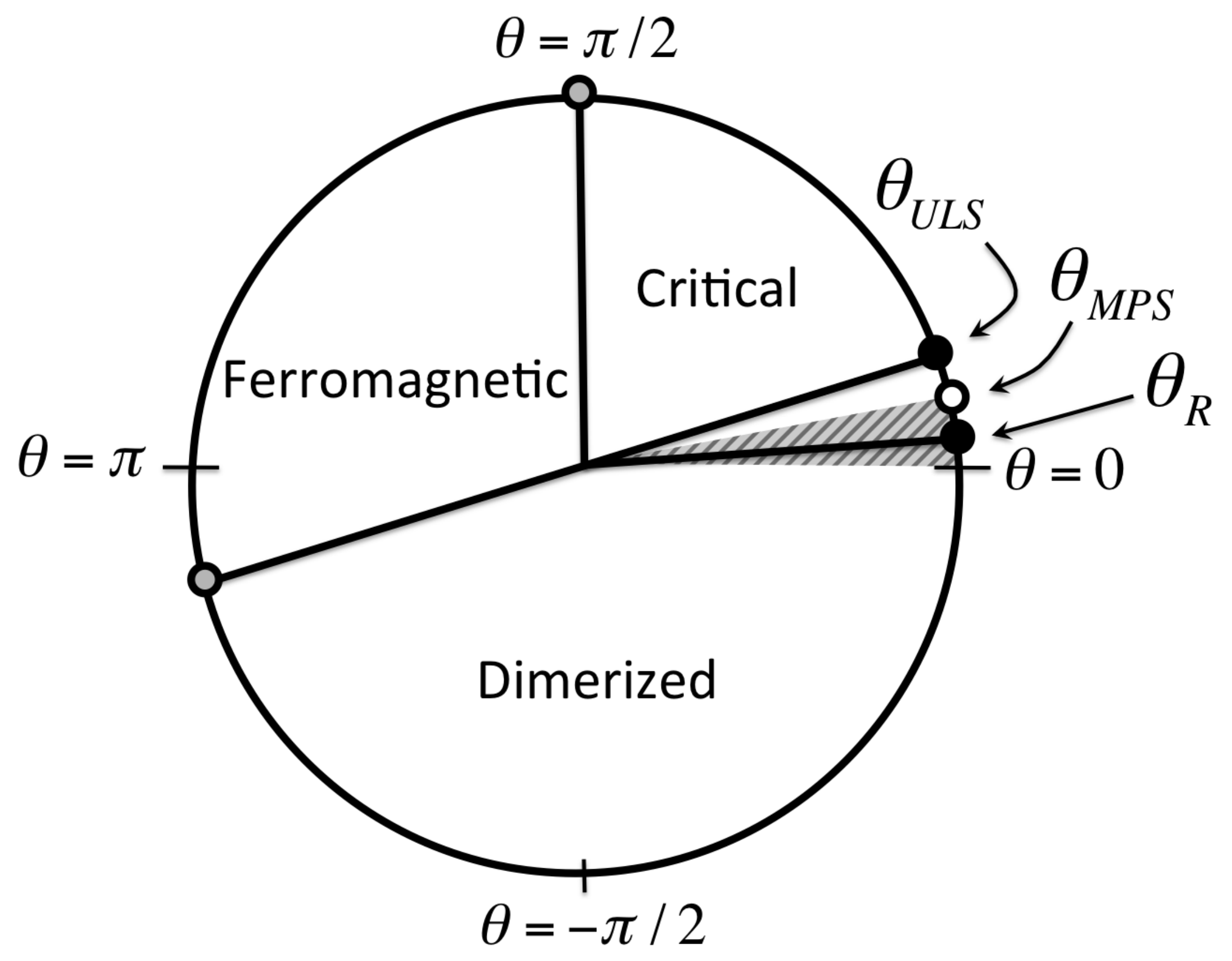}
\caption{Predicted phase diagram of the SO(6) bilinear-biquadratic spin
chain in Ref. \protect\cite{Tu-2008}. The shaded area is the region
considered in this Letter.}
\label{fig:SO(6)}
\end{figure}

Following Ref. \cite{Lecheminant-2004}, we derive the field theory for the
SO(6) Heisenberg chain by considering the SU(4) Hubbard model in the strong
coupling limit. Let us briefly review this approach to make the subsequent
discussions self-contained and to introduce some necessary notations. The
SU(4) Hubbard model is written as%
\begin{equation}
H_{\mathrm{SU(4)}}=-t\sum_{j,\alpha }(c_{j\alpha }^{\dagger }c_{j+1,\alpha }+%
\mathrm{H.c.})+U\sum_{j}(n_{j}-2)^{2},  \label{eq-Hubbard}
\end{equation}%
where $c_{j\alpha }^{\dagger }$ is the fermion creation operator at site $j$
with color index $\alpha =1,\ldots ,4$ and $n_{j}=\sum_{\alpha }c_{j\alpha
}^{\dagger }c_{j\alpha }$ is the fermion number operator. Here we consider $%
U>0$ and the case of a half-filled fermion energy band (two fermions per
site). For $U\gg t$, charge excitation is strongly suppressed due to a Mott
gap $\Delta _{c}\sim U$. Thus, two fermions frozen at each site constitute
six states $c_{\alpha }^{\dagger }c_{\beta }^{\dagger }|0\rangle $, which
belong to the vector representation of SO(6). In this limit, standard
perturbation theory in $t/U$ yields an SO(6) Heisenberg model \cite%
{Marston-1999}. With this correspondence in hand, the SO(6) generators $%
L^{ab}$ are written as $L^{ab}=\frac{1}{2}\sum_{\alpha \beta }c_{\alpha
}^{\dagger }T_{\alpha \beta }^{ab}c_{\beta }$, where the $4\times 4$ SU(4)
matrices $T^{ab}$\ are\ normalized as $\mathrm{Tr}(T^{ab}T^{cd})=4\delta
_{ac}\delta _{bd}$. Since SO(6) is a rank-3 algebra, we choose three
diagonal Cartan generators $L^{12}$, $L^{34}$, and $L^{56}$ with $T^{12}=\sigma
^{0}\otimes \sigma ^{z}$, $T^{34}=\sigma ^{z}\otimes \sigma ^{0}$, and $%
T^{56}=\sigma ^{z}\otimes \sigma ^{z}$, where $\sigma ^{0}$ and $\sigma ^{z}$
are $2\times 2$ identity and Pauli matrices, respectively.

The field theory for the SO(6) Heisenberg chain can be derived by applying
Abelian bosonization techniques to the SU(4) Hubbard model (\ref{eq-Hubbard}%
) \cite{Lecheminant-2004}. After linearizing the spectra around two Fermi
points $k_{F}=\pm 2\pi /a_{0}$, the fermion operators are decomposed into
left-moving and right-moving components as $c_{j\alpha }\rightarrow \sqrt{%
a_{0}}(\psi _{R\alpha }e^{ik_{F}x}+\psi _{L\alpha }e^{-ik_{F}x})$, where $%
a_{0}$ is the lattice spacing. These chiral fermions are related to boson
fields as $\psi _{R(L),\alpha }=(2\pi a_{0})^{-1/2}\zeta_{\alpha }\exp [\pm i%
\sqrt{\pi }(\phi _{\alpha }\mp \Theta _{\alpha })]$, where $\zeta_{\alpha }$
are Klein factors. The boson fields for U(1) charge and SO(6) spin channels
are just linear combinations of $\phi _{\alpha }$, defined by $\phi
_{c,(s_{1})}=(\phi _{1}\pm \phi _{2}+\phi _{3}\pm \phi _{4})/2$ and $\phi
_{s_{2},(s_{3})}=(\phi _{1}\pm \phi _{2}-\phi _{3}\mp \phi _{4})/2$ \cite%
{Azaria-2000,Lecheminant-2010}, respectively. Similar equations hold for their dual
fields. The bosonized Hamiltonian density $\mathcal{H}=\mathcal{H}_{0}+%
\mathcal{H}_{\mathrm{int}}$ contains the usual free part $\mathcal{H}_{0}=%
\frac{v_{c}}{2}[K_{c}(\partial _{x}\Theta _{c})^{2}+\frac{1}{K_{c}}(\partial
_{x}\phi _{c})^{2}]+\frac{v_{s}}{2}\sum_{p=1}^{3}[K_{s}(\partial _{x}\Theta
_{s_{p}})^{2}+\frac{1}{K_{s}}(\partial _{x}\phi _{s_{p}})^{2}]$ with
velocities $v_{c,(s)}$ and Luttinger parameters $K_{c,(s)}$.  The
interaction part is \cite{Lecheminant-2004} $\mathcal{H}_{\mathrm{int}%
}=g_{s}\sum_{p\neq q}\cos \sqrt{4\pi }\phi _{s_{p}}\cos \sqrt{4\pi }\phi
_{s_{q}}-g_{sc}\cos \sqrt{4\pi }\phi _{c}\sum_{p=1}^{3}\cos \sqrt{4\pi }\phi
_{s_{p}}$, where $g_{s},g_{sc}>0$ and the spin-charge coupling term comes
from the $4k_{F}$ umklapp scattering presence at half filling. Because of the
large Mott gap, the charge boson can be safely integrated out. By
introducing six Majorana fermions to refermionize the boson fields $\phi
_{s_{p}}$, for instance $(\xi ^{1}+i\xi ^{2})_{R(L)}\sim \exp [\pm i\sqrt{%
4\pi }\phi _{s_{1},R(L)}]$, the low-energy effective Hamiltonian density
describing the SO(6) spin sector reads \cite{Lecheminant-2004}%
\begin{eqnarray}
\mathcal{H}_{\mathrm{eff}} &=&-\frac{iv_{s}}{2}\sum_{\nu =1}^{6}(\xi
_{R}^{\nu }\partial _{x}\xi _{R}^{\nu }-\xi _{L}^{\nu }\partial _{x}\xi
_{L}^{\nu })-im\sum_{\nu =1}^{6}\xi _{R}^{\nu }\xi _{L}^{\nu }  \notag \\
&&-G_{s}\sum_{1\leq \nu <\kappa \leq 6}\xi _{R}^{\nu }\xi _{L}^{\nu }\xi
_{R}^{\kappa }\xi _{L}^{\kappa },  \label{eq-Majorana}
\end{eqnarray}%
with $m>0$ and $G_{s}>0$. Equation(\ref{eq-Majorana}) predicts a gapped
ground state with spin-Peierls order for the SO(6) Heisenberg chain \cite%
{Lecheminant-2004}, which was confirmed numerically \cite%
{Marston-1999}.

Let us now extend the effective theory in Eq.~(\ref{eq-Majorana}) for
the SO(6) Heisenberg chain in order to incorporate a
biquadratic interaction. In Eq.~(\ref{eq-Majorana}), the Majorana mass term
is relevant and the fermion interaction term is marginally irrelevant for $%
G_{s}>0$. The crucial observation here is that, when the Heisenberg chain is
perturbed by the biquadratic interaction, the \emph{only} permitted
form of the effective theory is still Eq. (\ref{eq-Majorana}) but with
modified parameters $v_{s}$, $m$, and $G_{s}$ as functions of $\theta $. The
reason for this is that in one spatial dimension only the fermion mass term
is relevant and the four-fermion interaction term is marginal \cite%
{Affleck-Haldane-1987}. Other terms are either irrelevant, which play no
role in the low-energy limit, or break the SO(6) symmetry, which
is forbidden since the effective theory must preserve the symmetry of the
original model.

More evidence for this effective theory comes from the integrable point $%
\theta _{\mathrm{R}}=\tan ^{-1}\frac{1}{8}$ for $n=6$. The Bethe ansatz
solution of this model, as shown by Minahan and Zarembo \cite{Minahan-2003},
indicates that there are three branches of gapless excitations above the
SO(6) singlet ground state. In the low-energy limit, these excitations have
the same linear dispersion $\varepsilon (p)\simeq v_{s}p$. This is in full
agreement with the effective theory (\ref{eq-Majorana}) with a vanishing
Majorana mass, which becomes the SO(6)$_{1}$ Wess-Zumino-Novikov-Witten theory (possibly perturbed
by marginally irrelevant terms) with central charge $c=6\times \frac{1%
}{2}=3$. At this critical point, the \textquotedblleft speed of
light\textquotedblright\ $v_{s}$\ in Eq. (\ref{eq-Majorana}) is given by $%
v_{s}=\frac{\pi }{2}\cos \theta _{\mathrm{R}}$, which is determined from the
Bethe ansatz solution. For $\theta <\theta _{\mathrm{R}}$, the Majorana mass
is expected to be decreasing when $\theta $ increases from $0$ to $\theta _{%
\mathrm{R}}$. For $\theta >\theta _{\mathrm{R}}$, it changes
sign ($m<0$) and an energy gap reopens.

Let us now derive the correlation functions. First, we note that the
effective theory in Eq.~(\ref{eq-Majorana}) is equivalent to six decoupled
Ising models and the Majorana mass is $m\sim (T-T_{c})/T_{c}$ with $T_{c}$
the Ising critical temperature \cite{Gogolin-1998}. For $\theta <\theta _{%
\mathrm{R}}$, the model in Eq.~(\ref{eq-BB}) is in the Ising disordered
phase with $\langle \mu _{\nu }\rangle \neq 0$ ($\nu =1,2,\ldots ,6$), while
for $\theta >\theta _{\mathrm{R}}$ it is in the Ising ordered phase with $%
\langle \sigma _{\nu }\rangle \neq 0$ (where $\mu _{\nu }$ and $\sigma _{\nu
}$ are the Ising disorder and order operators, respectively). In the continuum limit, the
SO(6) generators are expressed as $L_{j}^{ab}\simeq
J_{R}^{ab}+J_{L}^{ab}+(-1)^{j}n^{ab}$, where the slowly varying SO(6)
Kac-Moody currents $J_{R(L)}^{ab}$ and the staggered components $n^{ab}$
have critical dimensions $1$ and $3/4$, respectively. The $n^{ab}$'s
associated with the Cartan generators are $n^{12}\sim \sigma _{1}\sigma
_{2}\mu _{3}\mu _{4}\mu _{5}\mu _{6}$, $n^{34}\sim \mu _{1}\mu _{2}\sigma
_{3}\sigma _{4}\mu _{5}\mu _{6}$, and $n^{56}\sim \mu _{1}\mu _{2}\mu _{3}\mu
_{4}\sigma _{5}\sigma _{6}$. Using all this information, one can see that
the two-point correlator $\langle L_{j}^{ab}L_{j+r}^{ab}\rangle $ decays
exponentially in both Ising disordered and ordered phases, while at the
critical point $\theta _{\mathrm{R}}$ it decays algebraically as
\begin{equation}
\langle L_{j}^{ab}L_{j+r}^{ab}\rangle \simeq \frac{C_{1}}{r^{2}}+(-1)^{r}%
\frac{C_{2}}{r^{3/2}},
\end{equation}%
where $C_{1}$ and $C_{2}$ are nonuniversal constants. Furthermore, in Ref. \cite%
{Affleck-1991}, a dimerization operator $\mathcal{D}_{j}=(-1)^{j}%
\sum_{a<b}L_{j}^{ab}L_{j+1}^{ab}$ and an operator $\mathcal{C}%
_{j}=(-1)^{j}\sum_{a<b}L_{j}^{ab}L_{j+1}^{ab}L_{j+2}^{ab}$ characterizing
the staggered charge-conjugation order were suggested to distinguish between
the two phases. Remarkably, in the continuum limit we find that $\mathcal{D}%
\sim \mu _{1}\mu _{2}\mu _{3}\mu _{4}\mu _{5}\mu _{6}$ and $\mathcal{C}\sim
\sigma _{1}\sigma _{2}\sigma _{3}\sigma _{4}\sigma _{5}\sigma _{6}$,
satisfying $\langle \mathcal{D}_{i}\rangle \neq 0,\langle \mathcal{C}%
_{i}\rangle =0$ for $\theta <\theta _{\mathrm{R}}$ and $\langle \mathcal{D}%
_{i}\rangle =0,\langle \mathcal{C}_{i}\rangle \neq 0$ for $\theta >\theta _{%
\mathrm{R}}$, which indeed characterizes the two different phases. At the
critical point $\theta _{\mathrm{R}}$ these competing orders have the same
critical exponents and we have $\langle \mathcal{D}_{j}\mathcal{D}%
_{j+r}\rangle \sim 1/r^{3/2}$ and $\langle \mathcal{C}_{j}\mathcal{C}%
_{j+r}\rangle \sim 1/r^{3/2}$. Therefore, the critical ground state at $%
\theta _{\mathrm{R}}$ is an algebraic spin liquid unifying the dimerization
order, the staggered charge-conjugation order, and the N\'{e}el order.

For the solvable point $\theta_{\mathrm{MPS}}$, it was shown in Ref. \cite%
{Tu-2008}\ that the MPS has a hidden antiferromagnetic order quantified by a
generalized den Nijs-Rommelse string order parameter (SOP) \cite{den
Nijs-1989} $\mathcal{O}^{ab}=\lim_{|k-j|\rightarrow \infty }\langle
L_{j}^{ab}\prod_{l=j}^{k-1}\exp (i\pi L_{l}^{ab})L_{k}^{ab}\rangle $. Because of
the unbroken SO(6) symmetry, all these SOPs are equal and it is sufficient
to consider the SOPs for the Cartan generators. We find that they are
related to the Ising variables as%
\begin{equation}
\mathcal{O}^{12}\sim \langle \sigma _{1}\rangle \langle \sigma _{2}\rangle ,%
\text{ }\mathcal{O}^{34}\sim \langle \sigma _{3}\rangle \langle \sigma
_{4}\rangle ,\text{\ }\mathcal{O}^{56}\sim \langle \sigma _{5}\rangle
\langle \sigma _{6}\rangle   \label{eq:IsingSOP}
\end{equation}%
These SOPs have nonzero values in the Ising ordered phase for $\theta
>\theta _{\mathrm{R}}$ and vanish in the disordered phase for $\theta
<\theta _{\mathrm{R}}$. Therefore, they are also proper order parameters for
the phase with staggered charge-conjugation order and their utility goes beyond the solvable point $\theta _{\mathrm{MPS}}$.

\emph{Reduction from $n=6$ to $n=5$.-} In the following we establish a
relationship between the effective field theories for SO(6) and for SO(5)
\cite{Alet-2010}. Let us go back to the SU(4) Hubbard model in Eq.~(\ref%
{eq-Hubbard}) and interpret the fermion color index as spin-$\frac{3}{2}$
quantum numbers ($\alpha =\pm \frac{3}{2},\pm \frac{1}{2}$). Then, the SO(6)
vector representation at each site unifies the spin-$2$ quintet states and
the spin-$0$ singlet state formed by two spin-$\frac{3}{2}$ fermions. If an
on-site spin-dependent interaction $V\sum_{i}\mathbf{S}_{i}^{2}$\ is added
to the model (\ref{eq-Hubbard}), then the singlet and quintet sectors pick
up different energies and the SU(4) symmetry of the Hamiltonian is broken
down to SO(5) \cite{CWu-2006}, with an energy difference $\Delta _{s}\sim V$
between the two sectors. For $V<0$ the quintet states, forming SO(5) vector
representation, have lower energy and the effective exchange Hamiltonian in
the Mott regime is an SO(5) Heisenberg model \cite{CWu-2006,Tu-2006}. In the
field theory treatment \cite{Alet-2010}, the quintet and singlet degrees of
freedom are described by Majorana fermions $\xi ^{\nu }$ ($\nu =1\sim 5$)
and $\xi ^{6}$, respectively. In Eq. (\ref{eq-Majorana}), the energy
difference $\Delta _{s}$ is formally accounted for by giving $\xi ^{6}$ a
large mass $m_{\mathrm{s}}\gg m_{\mathrm{q}}>0$, where $m_{\mathrm{q}}$ is
the mass of the other five Majorana fermions. In the low-energy limit,
integration over $\xi ^{6}$ yields five massive Majorana fermions with a
renormalized mass $m_{\mathrm{q}}^{\prime }>0$. Although the five remaining
Majorana fermions have a \textit{smaller} mass $m_{\mathrm{q}}^{\prime }<m_{%
\mathrm{q}}$, the SO(5) Heisenberg chain is still in the Ising disordered
phase, corresponding to dimerized ground states \cite{Alet-2010}.

Thanks to this mechanism, we can now study the correlation functions of the
SO(5) model. This is accomplished by simply replacing the Ising variables $
\mu _{6}$ and $\sigma _{6}$ in the expressions for SO(6) by their expectation
values $\langle \mu_{6}\rangle \neq 0$ and $\langle \sigma_{6}\rangle =0$, respectively. In
the SO(5) Ising ordered phase, unlike the SO(6) case, translational symmetry
is preserved and all two-point correlation functions decay exponentially.
For example, it is easy to show that for SO(5) the operator $%
\mathcal{C}$ has a vanishing expectation value. However, according to Eq. (%
\ref{eq:IsingSOP}), the nonlocal SOPs are still nonzero and
thus are valid order parameters in this phase.

\emph{Reduction and effective theory for arbitrary $n$.-} Motivated by the
above discussions, we now propose a reduction mechanism to understand the
ground state of the SO($n$) Heisenberg model. As we have seen, the SO($n$)
Heisenberg models for $n=6$ and $5$ are both described by $n$ massive
Majorana fermions with mass $m>0$ and marginally irrelevant terms. Formally speaking, one can say that we go from an effective
theory for SO($n$) to an effective theory for SO($n-1$) by giving a large
mass to the Majorana fermion $\xi ^{n}$ and integrating it out. Because of
the marginally irrelevant couplings in the SO($n$) effective theory, the
elimination of the large-mass Majorana fermion induces a \textit{negative}
contribution to the mass of the $n-1$ remaining Majorana fermions. Usually,
the exact values of the parameters in these effective theories are not
known. However, the subtlety here is that the SO(4) Heisenberg chain lies
exactly at the critical point $\theta _{\mathrm{R}}$. Since SO(4)$\simeq $%
SU(2)$\times $SU(2), the SO(4) Heisenberg chain is equivalent to two
decoupled spin-$\frac{1}{2}$ Heisenberg chains, whose effective theory
contains four \textit{massless} Majorana fermions \cite{Nersesyan-1997}. Thus, if one wishes to construct
this theory from the effective theory for SO(5), the only possibility to
obtain the zero mass of the four Majorana fermions is the miraculous \textit{exact} \textit{cancellation} of their original mass [in the SO(5) theory]
and the negative contribution that comes from integrating over the
large-mass Majorana fermion $\xi ^{5}$. We can proceed further from this
massless SO(4) theory on to an SO(3) theory, where three massive
Majoranas with $m<0$ are obtained and which recovers Tsvelik's theory for the spin-$1$ Heisenberg chain \cite{Tsvelik-1990}.

What about the effective field theory for general $n$? By combining the
results for $n=3,\ldots ,6$ and applying the reduction mechanism in the
reverse direction, we can argue that at
the integrable point $\theta _{\mathrm{R}}$ the SO($n$) spin chain in Eq. (\ref%
{eq-BB}) is described by an SO($n$)$%
_{1}$ Wess-Zumino-Novikov-Witten model with $n$ massless Majorana fermions perturbed by marginally
irrelevant terms. This theory has central charge $c=n\times \frac{1}{2%
}$, and thus the entanglement entropy of a block increases linearly with $n$
for the ground state of the system close to and at $\theta _{\mathrm{R}}$
\cite{entang}. For $n\geq 5$, the isolated critical point $\theta _{\mathrm{R%
}}$ separates the region $\theta \in \lbrack 0,\theta _{\mathrm{MPS}}]$ into and
Ising disordered phase for $\theta <\theta _{\mathrm{R}}$ and an ordered phase
for $\theta >\theta _{\mathrm{R}}$. By using the results in Ref. \cite{Lecheminant-2002}, $n$ zero-energy Majorana modes $\eta^a$ ($a=1\sim n$) localized at the boundary are obtained from the SO($n$) effective theory for $m<0$ in a semi-infinite chain and form SO($n$) generators $\Gamma^{ab}=i\eta^{a}\eta^{b}$ in spinor representation. Note that the SO($n$) spinor is irreducible for odd $n$ but reducible for even $n$ and contains two subspaces. This explains the appearance of SO($n$) spinor edge states at $\theta _{\mathrm{MPS}}$ and their crucial differences for even and odd $n$ \cite{Tu-2008}. Another interesting prediction of our theory is the two-point correlator $\langle
L_{j}^{ab}L_{j+r}^{ab}\rangle $ at the critical point $\theta _{\mathrm{R}}$%
. In the continuum limit, the SO($n$) generators $L^{ab}$ are a sum of
uniform SO($n$) currents with critical dimension $1$ and staggered
components with critical dimension $n/8$ ($n$ Ising variables). For $3<n<7$,
the contribution from the staggered components is dominant in the correlator
and we have $\langle L_{j}^{ab}L_{j+r}^{ab}\rangle \sim (-1)^{r}/r^{n/4}$.
However, for $n>8$, the contribution from uniform SO($n$) currents becomes
dominant and in this case $\langle L_{j}^{ab}L_{j+r}^{ab}\rangle \sim 1/r^{2}
$.

H.H.T.\ is grateful to G.-M. Zhang and T. Xiang for earlier
collaborations, and acknowledges discussions with I. Cirac, M. Aguado, M. Cheng, H. Katsura, P. Lecheminant, Z.-X. Liu, and S. Rachel. R.O. acknowledges support from the EU.

\end{document}